\documentclass[aps,twocolumn,amsfonts,amssymb,longbibliography,superscriptaddress]{revtex4-1}
\usepackage{graphicx}
\usepackage{multirow}
\usepackage{epstopdf} 
\usepackage{dcolumn}
\usepackage{bm}
\usepackage{bbm,bbold}
\usepackage{color}
\usepackage{xcolor, soul}
\sethlcolor{green}
\usepackage{amsfonts}
\usepackage{amsmath}
\usepackage{mdframed}
\usepackage[normalem]{ulem}
\usepackage{mathrsfs}   
\usepackage[none]{hyphenat}
\usepackage{subfigure}
\usepackage{float}
\usepackage{physics}
\usepackage[colorlinks=true,citecolor=blue]{hyperref}
\hypersetup{colorlinks=true,citecolor=blue,linkcolor=blue,urlcolor=blue}

\begin{document}
	\title{Longitudinal DC Conductivity in Dirac Nodal Line Semimetals: Intrinsic and Extrinsic Contributions}	
	\author{Vivek Pandey}
    \email{vivek_pandey@srmap.edu.in}
	\affiliation{Department of Physics, School of Engineering and Sciences, SRM University AP, Amaravati, 522240, India}
	\author{Dayana Joy}
    \email{dayana_joy@srmap.edu.in}
	\affiliation{Department of Physics, School of Engineering and Sciences, SRM University AP, Amaravati, 522240, India}
	\author{Dimitrie Culcer}
    \email{d.culcer@unsw.edu.au}
	\affiliation{School of Physics, The University of New South Wales, Sydney 2052, Australia}
 	\author{Pankaj Bhalla}
	\email{pankaj.b@srmap.edu.in}
 \affiliation{Department of Physics, School of Engineering and Sciences, SRM University AP, Amaravati, 522240, India}
    
\date{\today}

\begin{abstract}
Nodal line semimetals, a class of topological quantum materials, exhibit a variety of novel phenomena due to their properties, such as bands touching on a one-dimensional line or a ring in the Brillouin zone and drumhead-like surface states. In addition, these semimetals are protected by the combined space-inversion and time-reversal ($\mathcal{PT}$) symmetry. In this study, we investigate the longitudinal DC conductivity of the Dirac nodal line semimetals for the broken $\mathcal{PT}$-symmetric system by the mass term. Here, using the quantum kinetic technique, we find the intrinsic (field-driven) and extrinsic (scattering-driven) contributions to the total DC conductivity due to interband effects.  Interestingly, the resulting intrinsic conductivity is the Fermi sea contribution, while the extrinsic stems from the Fermi surface contribution. We show that at low chemical potential, the extrinsic part contributes more and dominates over the traditional Drude intraband term, while at the high chemical potential, the intrinsic conductivity contributes. Furthermore, the total DC response due to interband effects saturates at high chemical potential and its strength decreases with increasing mass value. Our findings suggest that the extrinsic contributions are rich enough to understand the overall feature of the response for the three-dimensional system. 

\end{abstract}

\maketitle

\section{Introduction}
Since the development of topological quantum materials, they have become the center point for new-generation device fabrication due to their versatile applications such as topological field-effect transistors~\cite{cho_nl2011, zhu_sr2013}, topological p-n junctions~\cite{Habib_prl2015, eschbach_nc2015} and quantum computation~\cite{he_fop2019, Luo_NRP2022}.
These materials include topological insulators and topological semimetals~\cite{Nitesh_cr2021}. Topological insulators host a gapless surface state and a gapped bulk state, whereas the topological semimetals have the gapless bulk state~\cite{Baum_prl2015}. Furthermore, these topological semimetals are categorized into Dirac Semimetals (DSMs), such as Cd$_3$As$_2$~\cite{Wang_prb2013, liu_nm2014, Borisenko_prl2014, yi_sr2014}, Na$_3$Bi~\cite{liu_science2014, xu_science2015, Bernardo_am2021}, Weyl Semimetals (WSMs), including the TaAs family~\cite{Lv_prx2015, Weng_prx2015, xu_nc2016, Belopolski_prl2016, Souma_prb2016, Xu_sa2015}, WTe$_2$~\cite{soluyanov_nature2015, Wu_prb2016, li_nc2017} as well as Nodal Line Semimetals (NLSMs)~\cite{Burkov_prb2011} e.g. Cu$_3$N~\cite{Kim_prl2015}, Ca$_3$P$_2$~\cite{Chan_prb2016, Shuo_apx2018}, ZrSiS family~\cite{Fu_sa2019, Neupane_prb2016, Hu_prl2016, Takane_prb2017, Chen_prb2017}, PbTaSe$_2$~\cite{Bian_nature2016}, and BaNbS$_3$~\cite{Liang_prb2016} etc. 
DSMs host a singular band crossing point, which is fourfold degenerate~\cite{xie_s&a2020, Wang_prb2012, liu_science2014} and is protected by the time-reversal ($\mathcal{T}$) and space-inversion ($\mathcal{P}$) symmetry. Breaking one of these symmetries converts the singular band crossing point into pairs of singular points, which are twofold degenerate~\cite{xie_s&a2020, Lv_prx2015, yang_np2015}, and this creates WSMs~\cite{Burkov_prl2011}. In the case of the NLSMs, the conduction and valence bands touch each other in a one-dimensional line or loop in the Brillouin zone~\cite{Burkov_prb2011, Bian_nature2016, Shuo_apx2018, Bian_prb2016, Fang_prb2015, Yu_prl2015, Xie_APLM2015, Kim_prl2015, Ekahana_njp2017, Feng_prm2018, schoop2016dirac, Neupane_prb2016, Hu_prl2016, Takane_prb2017, Chen_prb2017, Wang_prb2017, Liang_prb2016}, which can host a higher concentration of Dirac and Weyl electrons~\cite{schoop_cm2018, singha_pnas2017}. 

\begin{figure}[htp]
    \centering
    \includegraphics[width=8cm]{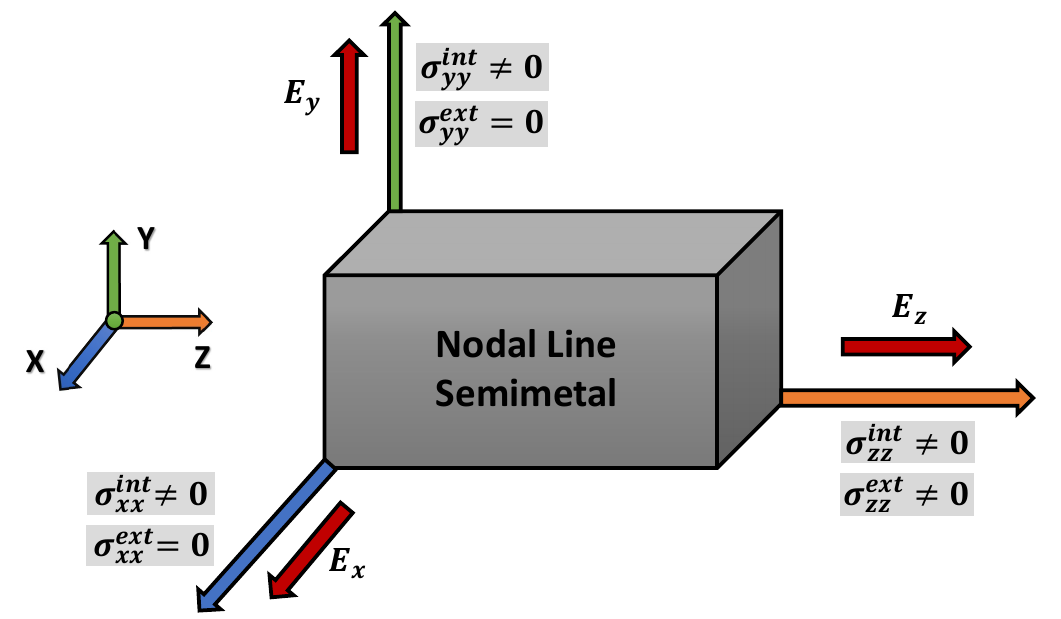}
    \caption{A schematic diagram depicts the longitudinal response of a $\mathcal{PT}$-symmetry broken three-dimensional Dirac Nodal Line Semimetal subjected to a static electric field along the $x$, $y$, and $z$ directions. In this setup, the sample exhibits intrinsic conductivity in all directions. However, the extrinsic conductivity contributes only along the axial direction ($z$-direction), attributed to the effects of disorder.}
    \label{fig:1.0}
\end{figure}
Based on band degeneracy, NLSMs can be divided into two baskets such as Dirac nodal line semimetals (DNLSMs)~\cite{Yu_prl2015, Xie_APLM2015, Feng_prm2018,schoop2016dirac,takane2018observation} and Weyl nodal line semimetals (WNLSMs)~\cite{Bian_nature2016}. Similar to Dirac and Weyl semimetals, DNLSMs are fourfold degenerate~\cite{Fang_prb2015, Kim_prl2015, Mullen_prl2015, Weng_prb2015, Burkov_prb2011,hu_s&a2021}, while WNLSMs are twofold degenerate~\cite{Feng_prl2019, deng_nc2019, Yang_prl2020}. However, symmetry plays a crucial role in their classification.
The DNLSMs are detected in the system with mirror symmetry, $\mathcal{P}$-symmetry, $\mathcal{T}$-symmetry, or combined $\mathcal{PT}$-symmetry~\cite{Guo_mf2022, Fang_prb2015, Kim_prl2015, Zhang_prb2017, barati_prb2017, Weng_prb2015, Mullen_prl2015}. Here Ca$_3$P$_2$, Cu$_3$N and ZrSiS are examples of $\mathcal{PT}$-symmetric DNLSMs~\cite{Xie_APLM2015, Chan_prb2016, Kim_prl2015, Fu_sa2019}. Breaking these symmetries can cause the degeneracy to vanish and may open a gap, which can result in different topological phases~\cite{Guo_mf2022, Weng_jopcm2016}. On the other hand, the WNLSMs lack symmetry, which can be noticed either in broken time-reversal symmetry or inversion symmetry lacking systems~\cite{Zhang_prb2020}. 
The presence of the mass term in DNLSMs breaks combined $\mathcal{PT}$-symmetry and opens up a gap in the band structure~\cite{wang_prb2021, Flores-Calderón_EL2023}. The mass term originates from factors such as electric fields, disorder, circularly polarized light, pressure, or uniaxial strain~\cite{chiba_prb2017, chen_prb2018, Kot_prb2020, wang_prb2021, Flores-Calderón_EL2023}. 

\begin{figure*}[htp]
    \centering
    \includegraphics[width=18cm]{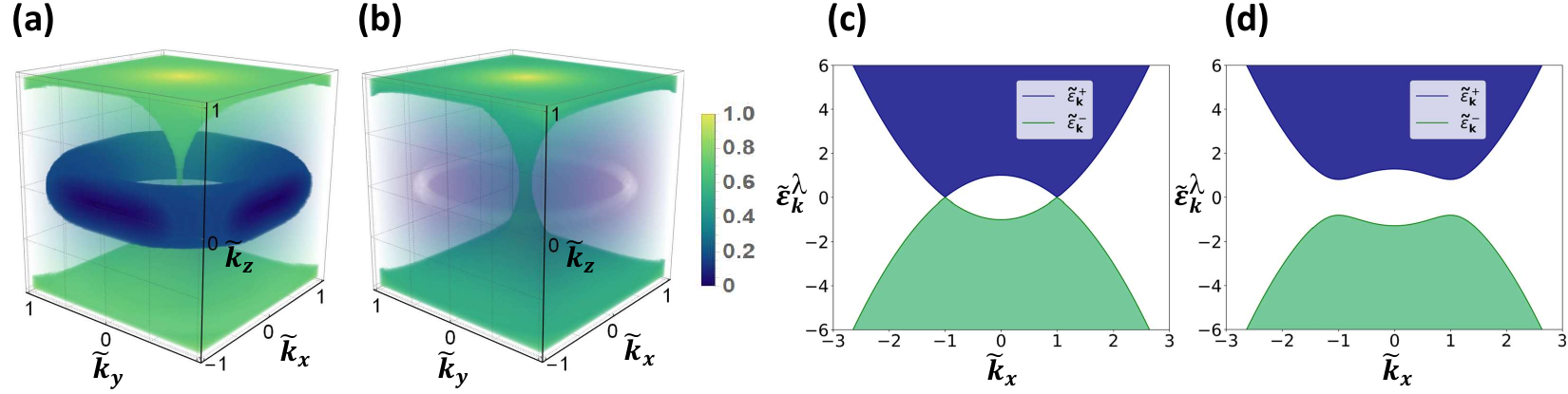}
    \caption{(a) and (b) 3D dispersion plot of DNLSMs with the mass term $\Tilde{M}=0$ where the $\mathcal{PT}-$symmetry remains intact and with the finite mass term $\Tilde{M}=0.8$ which breaks the $\mathcal{PT}-$symmetry of the system respectively. (c) and (d) 2D dispersion plot of DNLSMs with mass term $\Tilde{M}=0$ and $\Tilde{M}=0.8$.}
    \label{fig:2}
\end{figure*}

Due to the interesting band structure, NLSMs host numerous exotic transport properties. These include the planar Hall response in NLSMs resulting from the application of electric and magnetic fields in the same plane~\cite{Li_prb2023, Wang_prl2024}, the anomalous Hall effect in Heusler compounds originating from the Berry curvature effect~\cite{Chatterjee_jpcm2022, Chatterjee_prb2023}, the effect of the uniaxial strain on the intrinsic contribution of the anomalous Hall effect~\cite{Lim_prm2023}, and quantum transport phenomenon~\cite{zhao_qf2023, Yang_apx2022, kim_nc2022, Fu_njp2024, Rui_prb2018, Syzranov_prb2017}. The Drude conductivity (intraband) for DNLSMs has also been discussed in the DC regime~\cite{barati_prb2017,mukherjee_prb2017}.
Recently, Barati et al. studied the optical properties of two and three-dimensional nodal line semimetals~\cite{barati_prb2017}. Ref.~\cite{barati_prb2017} focussed on the intrinsic part of optical conductivity for the $\mathcal{PT}$-symmetric system by employing the Kubo formalism, and discussed the optical conductivity variation linearly with the chemical potential. On the other hand, Wang et al. calculated anomalous Hall AC and DC conductivity for the $\mathcal{PT}$-symmetry broken system by incorporating the tilt and the mass terms~\cite{wang_prb2021}. In Ref.~\cite{wang_prb2021} it was found that the nonzero Hall current coming from intrinsic contribution (field driving term) is generated at finite tilt and finite Fermi energy.
%
Although the Drude part dominates in the DC conductivity in various materials, the total response is the sum of both interband and intraband contributions. Further, the interband effects originate from distinct channels. One is from the field-driven mechanisms, also known as the intrinsic mechanism and the other is the interband scattering effects due to the intraband contributions, also known as the extrinsic contributions stemming from the intraband counterparts. In the literature, the main focus has been given to the intrinsic interband contribution and the intraband contribution (Drude) of NLSMs. However, the analysis is incomplete in estimating the strength of the net response of the system without taking into account other intraband-related effects, which can be larger than the Drude intraband part. This motivates the investigation of the interband effects which can boost the total signal of the system, and the explicit analysis of the extrinsic effects in the NLSMs in this work. Specifically, our research focuses on the longitudinal DC response in three-dimensional Dirac nodal line semimetals, considering the following aspects: 1: The impact of disorder (scattering effects) on the response of Dirac nodal line semimetals, 2: The influence of intraband elements of the density matrix on the interband scattering contribution, leading to the extrinsic component of DC conductivity, 3: The effect of the \(\mathcal{PT}\)-symmetry-broken mass term on longitudinal DC conductivity, 4: The treatment of the Fermi sea (intrinsic) and Fermi surface (extrinsic) contributions on equal footing.

In this article, we investigate the longitudinal conductivity in $\mathcal{PT}$-symmetric broken DNLSMs employing the quantum kinetic approach~\cite{culcer_prb2017, Bhalla_prb2023} and emphasize the interband intrinsic and extrinsic contributions to the longitudinal conductivity. The schematic picture of the different components of the longitudinal response to the distinct directions of the electric field is given in Fig.~\ref{fig:1.0}. We find that the extrinsic contribution stems from two different pathways: one is the interband scattering term due to the intraband contribution, which is the Fermi surface effect, and the other is the scattering term due to the interband driving term, which is the Fermi sea effect. However, the main extrinsic contribution is made by the Fermi surface part. We observe that the extrinsic effects provide a significant contribution to the longitudinal response as the intrinsic effect does, thereby enhancing the net signal of the response. Furthermore, our findings highlight the importance of disorder which will be relevant to interpret the experimental investigations. 

Our paper is organized as follows. In Section~\ref{sec:model}, we discuss the model Hamiltonian for the DNLSMs and provide the recipe for the quantum kinetic approach to obtain intrinsic and extrinsic contributions to the longitudinal conductivity. In Section~\ref{sec: results}, we present numerical results for DNLSMs. Finally, in Section~\ref{sec: summary}, we conclude by summarizing our work and providing future directions.

\section{Model and Quantum Kinetic Approach}\label{sec:model}
\subsection{Nodal Line Semimetal}
We consider a low-energy model Hamiltonian for the three-dimensional DNLSMs, a $\mathcal{PT}$-symmetry broken system~\cite{barati_prb2017, wang_prb2021, Flores-Calderón_EL2023}
\begin{equation}\label{eqn:hamitonian}
    \mathcal{H}({\bm k}) =  \frac{\hbar^{2}} {2m}\space(\mathcal{K}-k_0) \space\sigma_x +  \hbar\space v_z k_z \space\sigma_y + M \space \sigma_z. 
\end{equation}
Here, $\mathcal{K} = {\sqrt{k_x^2 + k_y^2}}$, $k_0$ is the radius of the nodal ring, $v_i$ and $k_i$ represents the velocity and the wave vector of an electron in the $i^{\rm th}$-direction respectively. Further, $\sigma_i$ refers to the Pauli matrix in pseudo spin basis having $i\equiv x, y, z$ and $M$ corresponds to the mass term. Based on the existing experimental and theoretical predictions, the mass term can be generated in various ways such as external electric field, uniaxial strain, pressure, disorder, circularly polarized light~\cite{chiba_prb2017, chen_prb2018, Kot_prb2020, wang_prb2021, Flores-Calderón_EL2023}. Furthermore the introduction of the mass term $M \sigma_z$ in the Hamiltonian breaks the $\mathcal{PT}$-symmetry of the system and generates a gap between the conduction and valence bands~\cite{chiba_prb2017, chen_prb2018, Kot_prb2020, wang_prb2021, Flores-Calderón_EL2023}. On the other hand in the absence of the mass term ($M=0$), $\mathcal{PT}$-symmetry of the system will remain intact. 
The energy eigenvalues for the Hamiltonian (Eq.~\eqref{eqn:hamitonian}) are
\begin{equation}
    \Tilde{\varepsilon}_{\bm k}^\lambda =  \lambda \sqrt{(\Tilde{\mathcal{K}} - 1)^2+(\gamma \Tilde{k}_z)^2 + \Tilde{M}^2},
\end{equation}
where we set, $ \Tilde{\varepsilon}_{\bm k}^\lambda = \varepsilon_{\bm k}^\lambda/\varepsilon_0, \Tilde{\mathcal{K}} =\mathcal{K}/k_0, \Tilde{k}_z=k_z/k_0, \gamma=2 m v_z/\hbar k_0$, $\Tilde{M}=M/\varepsilon_0$, $\varepsilon_0=\hbar^2 k_0^2/(2m_e)$ and $\lambda = +(-)$ corresponds to the conduction (valence) band.
The energy dispersion for this case is illustrated in Fig.~\ref{fig:2}. Here we can see that at mass $\Tilde{M}=0$ and $\Tilde{k}_z = 0$, the energy eigenvalues are degenerate and the dispersion shows the drumhead-like surface state as $\Tilde{\mathcal{K}}$ approaches unity~\cite{Rui_prb2018, Chan_prb2016,deng_nc2019}. Upon introducing the mass term the degeneracy of the system lifts up and a gap is created around the Dirac points. 
Further, the eigenfunctions associated with the Hamiltonian are
\begin{equation} \label{eqn:estate}
    | u_{\bm k}^{\lambda} \rangle =\sqrt{\frac{\Tilde{\varepsilon}_{\bm k}^\lambda+\Tilde{M}}{2 \Tilde{\varepsilon}_{\bm k}^\lambda}}  \left(\begin{matrix}
     -\sqrt{\frac{\Tilde{\varepsilon}_{\bm k}^\lambda-\Tilde{M}} {\Tilde{\varepsilon}_{\bm k}^\lambda +\Tilde{M}}} e^{-i\theta_k} \\
     {1}
    \end{matrix}\right),
\end{equation}
where $\theta_{\bm{k}}$ is the angle between the $\mathcal{K}$-plane and the $k_z$-direction and is defined as $\theta_{\bm{k}} = \tan^{-1} [\gamma \Tilde{\bm{k}}_z/ (\Tilde{\mathcal{K}}-1)]$.

According to the definition of the electrical current in response to the external field $\bm{j} = \text{Tr}\big[ \bm{v} \rho \big]$, where $\bm{v}$ is the velocity operator and $\rho$ is the density operator. This generates the component of the conductivity as $\sigma_{ij}=-e \sum_{\bm{k}} \big[\space v_i\space \rho_{{\bm k}}\big]/E_j$. Below, we discuss the quantum kinetic approach to compute the density matrix. 

\subsection{Quantum Kinetic Approach}
We first begin with the quantum Liouville equation for a single-particle density matrix averaged over impurity configurations~\cite{culcer_prb2017, Bhalla_prb2023}
\begin{equation} \label{eqn:QLE}
    \frac{\partial \rho}{\partial t}+\frac{i}{\hbar} [ \mathcal{H},\rho]=0. 
\end{equation}
Here, $\mathcal{H}= \mathcal{H}_0+\mathcal{H}_E+U$ is the total Hamiltonian of the system, and $[\cdot,\cdot]$ refers to the commutator bracket. In the former, $\mathcal{H}_0$ represents the unperturbed Hamiltonian or band Hamiltonian of a system, which yields $\mathcal{H}_0\ket{u_{\bm{k}}^n} = \varepsilon_{\bm{k}}^n\ket{u_{\bm{k}}^n}$, where $\ket{u_{\bm{k}}^n}$ is the Bloch wave function, having $\bm{k}$ as a wave vector and $n$ as a band index. The perturbed Hamiltonian $\mathcal{H}_E =e\bm{E}$.$\hat{\bm{r}}$ arises due to the interaction of an external electric field $\bm{E}$ with the sample. Here, we consider the static and spatially homogeneous electric field and $\hat{\bm{r}}$  as the position vector of an electron. The third part of the total Hamiltonian, $U$ refers to the disorder potential (here we have taken weak disorder potential). Inserting the full Hamiltonian in the kinetic equation Eq.~\eqref{eqn:QLE} brings the commutator between the Hamiltonian and the density matrix into three parts. The first part $[\mathcal{H}_0,\rho]$ corresponds to the interband transition term. The second part $[\mathcal{H}_E,\rho]$ refers to the driving term, where the driving term in the band basis representation is $\hbar D_{E,\bm{k}}^{np} =\hbar \langle n,{\bm {k}}|[\mathcal{H}_E,\rho]\ket{p,{\bm {k}}} = e\bm{E}.\big[\mathcal{D}_{\bm{k}}\rho\big]^{np}$,  where $\big[\mathcal{D}_{\bm{k}}\rho\big]^{np} = \partial_{\bm {k}} \rho^{np}-i[\mathcal{R}_{\bm{k}},\rho]^{np}$ is a covariant derivative~\cite{Nagaosa_AM2017, Bhalla_prb2023} and $\ket{n,{\bm{k}}}= e^{-i\bm{k.r}}\ket{u_{\bm{k}}^n}$ represents the eigenfunction of the system. $\mathcal{R}_{\bm{k}}^{np}=\bra{u_{\bm{k}}^n}\ket{i\partial_{\bm{k}}u_{\bm{k}}^p}$ is the Berry connection in the $\bm{k}$-space and $\partial_{\bm{k}}$ is the partial derivative with respect to the wave vector. The third $\mathcal{J}[\rho] =[U,\rho]$ is the scattering term, which we treat within the first-order Born approximation~\cite{culcer_prb2017}. However, this approximation will no longer will valid in the case of strong disorder potential. The former scattering term in the band basis representation is defined in the form
\begin{align}\nonumber\label{eqn:J}
\mathcal{J}_{\bm{k}}^{np}[\rho]=\sum_{\bm{k}'}\sum_{qt}\bigg[\frac{U_{\bm{kk}'}^{nq}U_{\bm{k}'\bm{k}}^{qt}\space\rho^{qp}_{\bm{k}}}{i\big(\varepsilon_{\bm{k}'}^q-\varepsilon_{\bm{k}}^t\big)/\hbar}-\frac{U_{\bm{kk}'}^{nq}U_{\bm{k}'\bm{k}}^{tp}\space\rho^{qt}_{\bm{k}'}}{i\big(\varepsilon_{\bm{k}'}^t-\varepsilon_{\bm{k}}^p\big)/\hbar}\\
-\frac{U_{\bm{kk}'}^{nq}U_{\bm{k}'\bm{k}}^{tp}\space\rho^{qt}_{\bm{k}'}}{i\big(\varepsilon_{\bm{k}'}^n-\varepsilon_{\bm{k}}^q\big)/\hbar}+\frac{U_{\bm{kk}'}^{qt}U_{\bm{k}'\bm{k}}^{tp}\space\rho^{nq}_{\bm{k}}}{i\big(\varepsilon_{\bm{k}'}^q-\varepsilon_{\bm{k}}^t\big)/\hbar}\bigg].
\end{align}
Here, $U_{\bm{kk}'}^{nq} = \langle n,{\bm k} |U|q,{\bm k}\rangle$ is the disorder potential matrix. It is important to highlight that our study focuses on the disorder potential, with its spatial correlation function defined as follows: the average of the disorder matrix element is $\langle U(\bm {r})\rangle = 0$ and the multiple averages of disorder matrix elements are $\langle U(\bm{r})U(\bm{r}')\rangle=\frac{U_0^2}{V}\delta(\bm{r}-\bm{r}')$. Here, $U_0$ is the parameter that incorporates the strength of the disorder potential and $V$ as a unit volume. This results in the product of disorder potential in $\bm{k}$-space taking the form $U_{\bm{kk}'}^{nq}U_{\bm{k}'\bm{k}}^{qt}= U_0^2 \langle u_{\bm k}^{n} | u_{{\bm k}'}^{q} \rangle \langle u_{{\bm k}'}^{q} | u_{{\bm k}}^{t} \rangle$ and the details for the DNLSMs are given in Appendix:~\ref{Appendix:A}.

To study the interband and intraband effects of a system, we can express the density matrix as follows:
 \begin{align} \label{eqn:dmatrix}
     \rho_{\bm{k}} = \rho_{0,\bm{k}}^{nn} + \mathcal{N}_{E,\bm{k}}^{nn}+S_{E,\bm{k}}^{np},  
 \end{align}
where $\rho_{0,\bm{k}}^{nn}$ denotes the equilibrium Fermi-Dirac distribution function and is defined as $\rho_{0,\bm{k}}^{nn}=f^0{(\varepsilon_{\bm{k}})}=\big[1+e^{\beta(\varepsilon_{\bm{k}}-\mu)}\big]^{-1}$, having $\beta = [k_BT]^{-1}$, $k_B$ represents the Boltzmann constant, $T$ denotes the absolute electronic temperature and $\mu$ is the chemical potential. The last two terms in Eq.~\eqref{eqn:dmatrix} 

$\mathcal{N}_{E,\bm{k}}^{nn}$ and $S_{E,\bm{k}}^{np}$, represent the intraband (diagonal) and interband (off-diagonal) parts of the density matrix respectively. 
Solving the Eq.~\eqref{eqn:QLE} for the diagonal part of the density matrix gives $\mathcal{N}_{E,{\bm{k}}}^{nn}=\frac{e\space\bm{E}.\hat{\bm{k}} \tau_{ia}^n}{\hbar}\frac{\partial \varepsilon_{\bm {k}}^{n}}{\partial k}\frac{\partial f_0(\varepsilon_{\bm{k}}^{n})}{\partial \varepsilon_{\bm{k}}^n}$, where $\frac{1}{\tau^\pm_{ia}}=\frac{n_i\space U_0^2}{\hbar}\int dk'\space k'\space\delta(\varepsilon_{\bm {k}'}^\pm - \varepsilon_{\bm {k}}^\pm)$ denotes the intraband relaxation time in the presence of short-range impurities~\cite{culcer_prb2017}. Similarly, for the off-diagonal part of the density matrix, we replace $\rho$ in Eq.~\eqref{eqn:QLE} with $S_{E,{\bm k}}$. This reduces the expression as
\begin{equation} \label{eqn:7}
    \frac{\partial  S_{E,{\bm{k}}}^{np}}{\partial t}+\frac{i}{\hbar} [ \mathcal{H}_0, S_{E,{\bm{k}}}^{np}] +\mathcal{J}_{\bm{k}}^{np}[\mathcal{N}_{E,{\bm k}}] + \frac{S_{E,{\bm{k}}}^{np}}{\tau_{ie}}={D_{E,{\bm{k}}}^{np}}. 
\end{equation}
Here ${D_{E,{\bm{k}}}^{np}} = - \frac{i}{\hbar} \langle n |[ \mathcal{H}_E, \rho_{0,\bm{k}}]|p\rangle=\frac{ie{\bm E}}{\hbar}\cdot\mathcal{R}_{\bm{k}}^{np} \big[f_0(\varepsilon_{\bm{k}}^n) - f_0(\varepsilon_{\bm{k}}^p)\big]$ stems from the off-diagonal part of the driving term and is termed as the intrinsic interband contribution, $\mathcal{J}_{\bm{k}}^{np}[\mathcal{N}_{E,{\bm k}}] =  \frac{i}{\hbar} \langle n| [U,\mathcal{N}_{E,{\bm k}}]|p\rangle$ is the scattering term which includes the disorder potential (i.e. $U$) which gives the extrinsic interband contribution to the conductivity. Here we have taken the weak disorder potential and treated $\mathcal{J}_{\bm{k}}^{np}[\mathcal{N}_{E,{\bm k}}]$ within the first-order Born approximation. On the other hand, we consider $\mathcal{J}_{\bm{k}}^{np}[ S_{E,{\bm{k}}}] =  S_{E,{\bm{k}}}/\tau_{ie}$ using the relaxation time approximation,  where $\tau_{ie}$ refers to the interband relaxation time.
In general, the scattering term comprises four terms based on the combination of diagonal and off-diagonal parts. The first two come from the contribution of the diagonal part of the density matrix to the scattering term, which gives $\mathcal{J}_{\bm k}^{nn}[\mathcal{N}_{E}]$ and $\mathcal{J}_{\bm k}^{np}[\mathcal{N}_{E}]$. The other two stem from the off-diagonal density matrix to the scattering term $\mathcal{J}_{\bm k}^{nn}[S_{E}]$ and $\mathcal{J}_{\bm k}^{np}[S_{E}]$. Here, we are focusing on the off-diagonal or interband part of the density matrix relying on the intraband contribution, i.e. $\mathcal{J}_{\bm k}^{np}[\mathcal{N}_{E}]$ which is a Fermi surface term.

On solving Eq.~\eqref{eqn:7}, we obtain the solution of the first-order differential equation as \begin{align}\label{eqn:S_E}
    S_{E,{\bm{k}}}^{np}= \frac{1}{g+i\hbar\space\omega^{np}}\bigg[\underbrace{{D_{E,{\bm{k}}}^{np}}}_{\text{Intrinsic}} + \underbrace{\mathcal{J}_{\bm{k}}^{np}[\mathcal{N}_{E,{\bm{k}}}]}_{\text{Extrinsic}}\bigg],
\end{align}
where $g=\frac{\hbar}{\tau_{ie}}$ represents the interband scattering energy scale, $ \hbar\omega^{np}=\big(\varepsilon_{\bm{k}}^n -\varepsilon_{\bm{k}}^p\big)$ is the energy difference between $n$ and $p$ energy levels.
%

The present approach is general and can be applied to a variety of systems. In this study, we consider the case of the NLSMs and investigate the response of the system due to the external electric field. Our calculations for the conductivity which incorporates the weak disorder effect are applicable in the limit $\mu \tau/\hbar \gg 1$. 

In this study, we mainly compute the longitudinal $\sigma_{ii}$ component due to the field along the $i$-direction, as other components for the considered system vanish. The detailed proof for the latter case is provided in Appendix~\ref{Appendix:B}. 

\subsection{Intrinsic contribution to the longitudinal conductivity }
The intrinsic contribution to the longitudinal conductivity stems purely from the driving term $D_{E,{\bm{k}}}^{np}$. Following Eq.~\eqref{eqn:S_E} and the definition for the conductivity, we obtain 
\begin{align}\label{eqn:sigma_zz}
\sigma_{ii}^{\text{int}} = \frac{e^2}{\hbar} \sum_{n\neq p}\sum_{ \Tilde{\bm k}} \frac{F^{np} \Tilde{\omega}^{pn}}{\big\{\Tilde{g}+\space i\space \Tilde{\omega}^{np}\big\}} \mathcal{\Tilde{R}}_{\bm{k}_i}^{np}\mathcal{\Tilde{R}}_{\bm{k}_i}^{pn}.
\end{align}
where $\Tilde{g}=\frac{\hbar}{\varepsilon_0 \tau_{ie}}$ is the interband scattering energy scale, $\Tilde{\omega}^{np}=\hbar \omega^{np}/\varepsilon_0$, and $F^{np}=f^0{(\Tilde{\varepsilon}_{\bm{k}}^n)}-f^0{(\Tilde{\varepsilon}_{\bm{k}}^p)}$ refers to the occupation probability between two separate bands.
Notice that in general, the velocity matrix is defined as $ \space \Tilde{v}^{pn}_i=\space\delta_{pn}\space\partial_{\Tilde{k}_i}\Tilde{\varepsilon}_{\bm{k}}^n+i\space\mathcal{\Tilde{R}}^{pn}_{k_i}\space\Tilde{\omega}^{pn}$. Here, the first term refers to the intraband contribution and the second term to the interband contribution of the Bloch velocity. In this case, only the interband velocity matrix element plays a role in the longitudinal conductivity due to interband effects. Additionally, the anomalous part of velocity here is zero as we are concerned with the longitudinal response.

Taking into account the DNLSMs system, the real part of the intrinsic longitudinal conductivity $\sigma_{zz}^{int}$ reduces to: 

\begin{align}\nonumber \label{eqn:sigintzz}
    &\sigma_{zz}^{\text{int}} = \sigma_0  \int \Tilde{\mathcal{K}}\space d\Tilde{\mathcal{K}}\space d\Tilde{k}_z \frac{F^{+-}}{\big(\Tilde{g}^2+ 4\Tilde{\varepsilon}_{\bm k}^2)} \frac{\Tilde{g}}{\Tilde{\varepsilon}_{\bm k}\big(\Tilde{\varepsilon}_{\bm k}^2-\Tilde{M}^2\big)}\\
    &\times\bigg\{\big(\Tilde{\mathcal{K}}-1\big)^2+\frac{\Tilde{M}^2 \gamma^2 \Tilde{k}_z^2}{\Tilde{\varepsilon}_{\bm k}^2}\bigg\},
\end{align}
where $\sigma_0 = \frac{e^2\gamma^2}{\hbar (2\pi)^2}$. Here, $\sigma_{xx}^{\text{int}}=\sigma_{yy}^{\text{int}}=\sigma_{zz}^{\text{int}}/2$ due to the angular variation of the integrand~\cite{Daniel_prb2020}. 
While expressing Eq.~\eqref{eqn:sigintzz}, we use the relation $\frac{1}{\Tilde{g}+\space i\space \Tilde{\omega}^{np}} = \frac{\Tilde{g}- \space i\space \Tilde{\omega}^{np}}{\Tilde{g}^2+\space 4\space (\Tilde{\omega}^{np})^2} $ to obtain real part of the intrinsic conductivity.
From Eq.\eqref{eqn:sigintzz}, it's evident that the real part of the intrinsic conductivity is directly related to the factor  $\Tilde{g}$, means to the interband relaxation time, denoted as $\tau_{ie}$. Therefore, the behavior of $\sigma_{zz}^{\text{int}}$ relies on different limiting regimes of $\Tilde{g}$ and $\Tilde{\mu}$. 
In the limit $\Tilde{\mu} \ll \Tilde{g}$ or $\mu \ll \frac{\hbar}{\tau_{ie}}$, the intrinsic conductivity follows $\sigma_{zz}^{\text{int}}\propto1/\Tilde{g}\propto \tau_{ie}$. Therefore, the intrinsic conductivity is finite only in the dirty limit ($\frac{1}{\tau_{ie}}\neq 0$) and diverges in the clean limit ($\frac{1}{\tau_{ie}}\rightarrow 0$) which is unphysical (from the mathematical point of view, all regimes are explored).
On the other hand, if $\Tilde{\mu} \gg \Tilde{g}$, we find $\sigma_{zz}^{\text{int}}\propto \Tilde{g} \propto 1/\tau_{ie}$. The latter part vanishes in the clean limit and yields a non-zero value in the dirty limit. 
Moreover, the leading-order term of the intrinsic conductivity with mass depends on $\Tilde{M}^{-2}$. Thus, the increase in mass or gap will cause the intrinsic conductivity to decrease.
\begin{table}[t]
    \begin{center}
    \fontsize{7pt}{7pt}\selectfont  
    \renewcommand{\arraystretch}{1.5}  
    \setlength{\arrayrulewidth}{1.2pt} 
    \begin{tabular}{|>{\centering\arraybackslash}p{0.7cm}|>{\centering\arraybackslash}p{2.0cm}|>{\centering\arraybackslash}p{1.5cm}|>{\centering\arraybackslash}p{1.5cm}|>{\centering\arraybackslash}p{1.5cm}|} \hline  
        
       \multirow{2}{*}{\textbf{S.No.}}& \multirow{2}{*}{\textbf{Regimes}} & \multicolumn{3}{c|}{\textbf{Power Law Dependence}}\\ \cline{3-5}
        & & \multicolumn{1}{c|}{\textbf{\multirow{2}{*}{\textbf{Mass ($\Tilde{M}$)}}}} & \textbf{Chemical Potential ($\Tilde{\mu}$)} & \textbf{Interband Scattering Energy Scale ($\Tilde{g}$)}\\ \hline  
        
        1&  $\Tilde{g}\ll\Tilde{\mu}\ll\Tilde{M}$& $\Tilde{M}^0$& $\Tilde{\mu}^0$& $\Tilde{g}^0$\\ 
        
        2&  $\Tilde{g}\ll\Tilde{M}\ll\Tilde{\mu}$& $\Tilde{M}^1$&$\Tilde{\mu}^{-1}$& $\Tilde{g}^0$\\  
        
        3&$\Tilde{g}\gg \Tilde{\mu}\gg\Tilde{M}$& $\Tilde{M}^0$&   $\Tilde{\mu}^{-1}$& $\Tilde{g}^{-1}$\\  
        
        4& $\Tilde{g}\gg\Tilde{M}\gg\Tilde{\mu}$& $\Tilde{M}^{-1}$&  $\Tilde{\mu}^0$&$\Tilde{g}^{-1}$\\   
        
         5&  $\Tilde{\mu}\gg\Tilde{g}\gg\Tilde{M}$&$\Tilde{M}^0$&$\Tilde{\mu}^{-1}$&$\Tilde{g}^{1}$\\  
        
        6& $\Tilde{M}\gg\Tilde{g}\gg\Tilde{\mu}$& $\Tilde{M}^0$& $\Tilde{\mu}^0$& $\Tilde{g}^{-2}$\\ \hline 
    \end{tabular}
    \caption{The table shows the power-law dependence of the extrinsic conductivity on quantities such as mass ($\Tilde{M}$), chemical potential ($\Tilde{\mu}$), and scattering energy scale ($\Tilde{g}$) in distinct regimes of interest.}
    \label{tab:SP}
\end{center}
\end{table}
\subsection{Extrinsic contribution to the longitudinal conductivity}
The scattering term $\mathcal{J}_{\bm{k}}^{np}[\mathcal{N}_E]$ leads to the extrinsic contribution to the conductivity. In this case, the longitudinal conductivity, following Eq.~\eqref{eqn:S_E} becomes: 
\begin{align}
\label{eqn:sigma_ext}
\sigma_{ii}^{\text{ext}} = - \frac{ie}{\hbar} \sum_{n\neq p}\bigg\{\Tilde{\mathcal{R}}_{k_i}^{np}\frac{\mathcal{J}_{\bm{k}}^{np}[\mathcal{N}_E]\hbar\Tilde{\omega}^{pn}}{\Tilde{g}+i\Tilde{\omega}^{np}}\bigg\}.
\end{align}
Further using Eq.~\eqref{eqn:J}, here the interband scattering term originating from the intraband part of the density matrix can be written in the form
\begin{align}\nonumber\label{eq:jn}
    &\mathcal{J}^{np}_{\bm{k}} [\mathcal{N}_E]= \frac{\pi\space n_i}{\hbar} \sum_{q~\bm{k}'}{U}_{\bm{k}'\bm{k}}^{np}\space {U}_{\bm{k}'\bm{k}}^{pn}\bigg[\big(\mathcal{N}_{E,{\bm {k}}}^{nn}-\mathcal{N}_{E,{\bm {k}'}}^{pp}\big)\space\delta{\big(\Tilde{\varepsilon}_{\bm {k}}^n - \Tilde{\varepsilon}_{\bm {k}'}^p\big)}\\
    & +\big(\mathcal{N}_{E,{\bm {k}}}^{qq}-\mathcal{N}_{E,{\bm {k}'}}^{pp}\big)\space\delta{\big(\Tilde{\varepsilon}_{\bm {k}}^q - \Tilde{\varepsilon}_{\bm {k}'}^p\big)}\bigg],
\end{align}
where ${U}_{\bm{k}'\bm{k}}^{np}\space {U}_{\bm{k}'\bm{k}}^{pn}$ refers to the product of elements of the scattering potentials and the form of the latter for the three-dimensional DNLSM system is given in Appendix~\ref{Appendix:A}.
\begin{figure}[t]
    \centering
    \includegraphics[width=8.75cm]{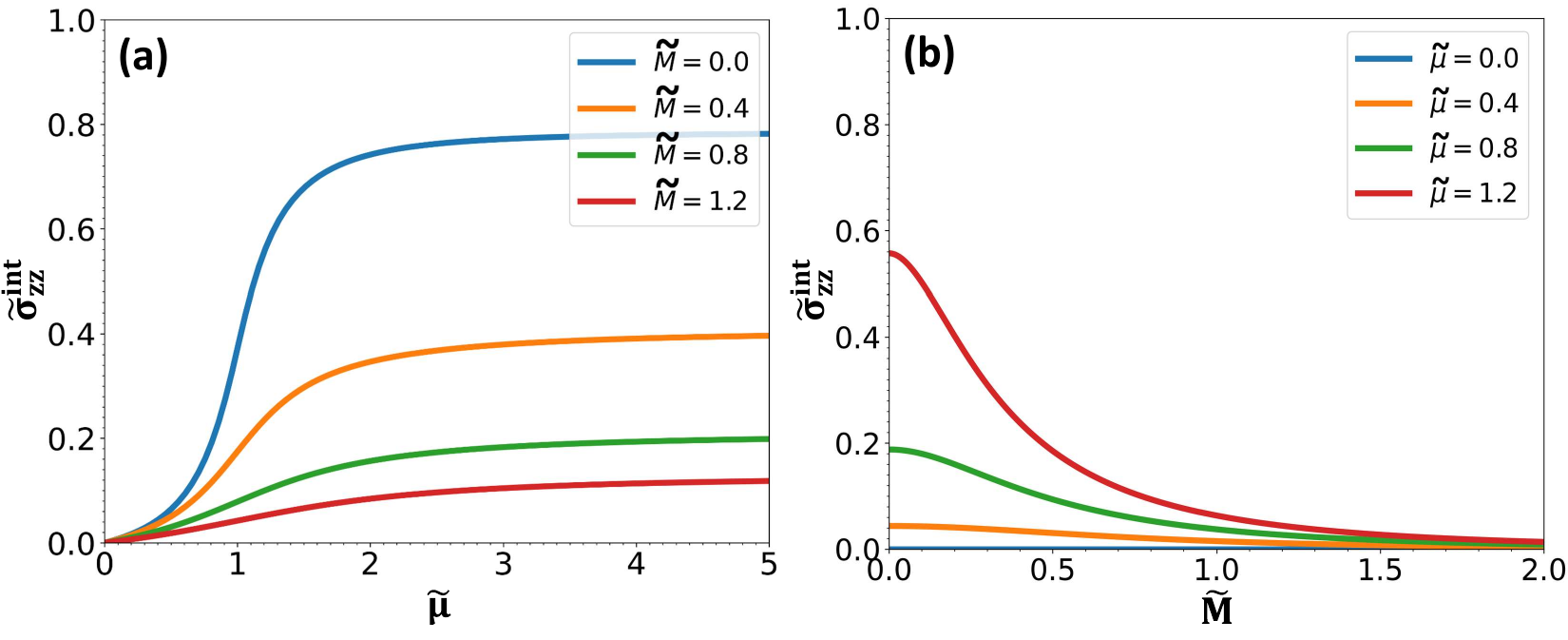}
    \caption{Plot of the normalized intrinsic longitudinal response in the $z$ -direction to the electric field response in the $z$-direction, i.e., $\Tilde{\sigma}_{zz}^{\text{int}}$= $\sigma_{zz}^{\text{int}}/\sigma_0$ (a) as a function of the normalized chemical potential at different mass values and (b) versus normalized mass at the different chemical potential. Here we consider $\Tilde{\mu}=\mu/\epsilon_0$ and $\Tilde{M}= M/\varepsilon_0$, $\Tilde{g}=\hbar/\varepsilon_0 \tau$ and set $\tau_{ie} =1 $ps.} 
    \label{fig:2.1}
\end{figure}
Corresponding to the Hamiltonian in Eq.~\eqref{eqn:hamitonian}, the scattering term on taking the field along the axial direction ($z$-direction), reduces to:
\begin{align}\nonumber\label{eqn:j_tau}
    &\mathcal{J}^{+-}_{\bm{k}} [\mathcal{N}_E]= \frac{i\pi n_i U_0^2}{2} \frac{e E_z}{\hbar^2} \sum_{\bm{k}'}\sin{(\theta_{k}'-\theta_k)} \frac{\sqrt{\Tilde{\varepsilon}^2_{\bm{k}'}}-\Tilde{M}^2}{\Tilde{\varepsilon}_{\bm{k}'}}\\[2ex]\nonumber 
    &\times\bigg\{-\tau_{ia}^- \space\frac{\partial{\Tilde{\varepsilon}_{\bm {k}'}^-}} {\partial{\bm {k}'}}\space{\delta(\Tilde{\varepsilon}_{\bm {k}'}^- - \Tilde{\mu})}\big[\space\delta(\Tilde{\varepsilon}_{\bm {k}}^+ - \Tilde{\varepsilon}_{\bm {k}'}^-)+ \delta(\Tilde{\varepsilon}_{\bm {k}}^- - \Tilde{\varepsilon}_{\bm {k}'}^-)\space\big]\\
    & +{\tau_{ia}^+ \space\frac{\partial{\Tilde{\varepsilon}_{\bm {k}'}^+}} {\partial{{\bm k}'}}\space{\delta(\Tilde{\varepsilon}_{\bm {k}'}^+ - \Tilde{\mu} )}\big[\space\delta(\Tilde{\varepsilon}_{\bm {k}}^+ - \Tilde{\varepsilon}_{\bm {k}'}^+ )+ \delta(\Tilde{\varepsilon}_{\bm {k}}^- - \Tilde{\varepsilon}_{\bm {k}'}^+)}\space\big]\bigg\},
\end{align}
where $n_i$ is the impurity density, $(\tau^{\pm}_{ia})^{-1}$ is the intraband relaxation time for the DNLSM $(\tau^{\pm}_{ia})^{-1}=\frac{n_i\space U_0^2 }{8\space\hbar \gamma \varepsilon_{\bm{k}}}$. The latter part, on inserting in the scattering term expression cancels the impurity density and $U_0^2$ dependence. On further algebraic calculations, 
the real part of the extrinsic conductivity $\sigma_{{zz}}^{\text{ext}}$ using Eq.~\eqref{eqn:sigma_ext} in response to the field along the $z$-direction becomes:  
\begin{align}\nonumber\label{eqn:sigextzz}
    &\sigma_{zz}^{\text{ext}}= \sigma_0 \frac{8\gamma}{\pi}\frac{1}{\space\big(\Tilde{\mu}^2-\Tilde{M}^2\big)}
\\\nonumber
&\times\int \Tilde{\mathcal{K}}'d\Tilde{\mathcal{K}}' \Tilde{\mathcal{K}} d\Tilde{\mathcal{K}} d\Tilde{k}_z \frac{\sqrt{\Tilde{\mu}^2-(\Tilde{\mathcal{K}}'-1)^2-\Tilde{M}^2}}{{\big(\Tilde{g}^2 + 4\Tilde{\varepsilon}_{\bm k}^2 \big)}}\\
&\times\bigg\{\Tilde{g}  (\Tilde{\mathcal{K}} -1)^2\space  + 2 \gamma \Tilde{M} (\Tilde{\mathcal{K}}-1) \Tilde{k}_z\space \bigg\}\delta(\Tilde{\varepsilon}_{\bm {k}}^+ - \Tilde{\mu}).
\end{align} 
Here, the real part of the extrinsic conductivity due to interband effects arises from the combination of the real and imaginary parts of various factors such as $\mathcal{\Tilde{R}}_{\bm{k}_i}^{np}, \mathcal{J}^{np}_{\bm{k}} [\mathcal{N}_E]$ and $\frac{\Tilde{g}-i\Tilde{\omega}^{np})}{\Tilde{g}^2+\space 4\space (\Tilde{\omega}^{np})^2}$. Interestingly, the above expression contains only the interband relaxation time scale in the form of $\Tilde{g}$.
Note that the above expression is expressed after performing the integration over angles associated with $\mathcal{K}$-plane and $\Tilde{k}_z^{'}$.
The variation of the above-complicated expression with the relaxation time, mass, and chemical potential in different regimes is discussed in Table~\ref{tab:SP}. In addition, the extrinsic contribution for other conductivity components vanishes, and the details are provided in Appendix~\ref{Appendix:B2}.  
\begin{figure}[t]
    \centering
    \includegraphics[width=8.75cm]{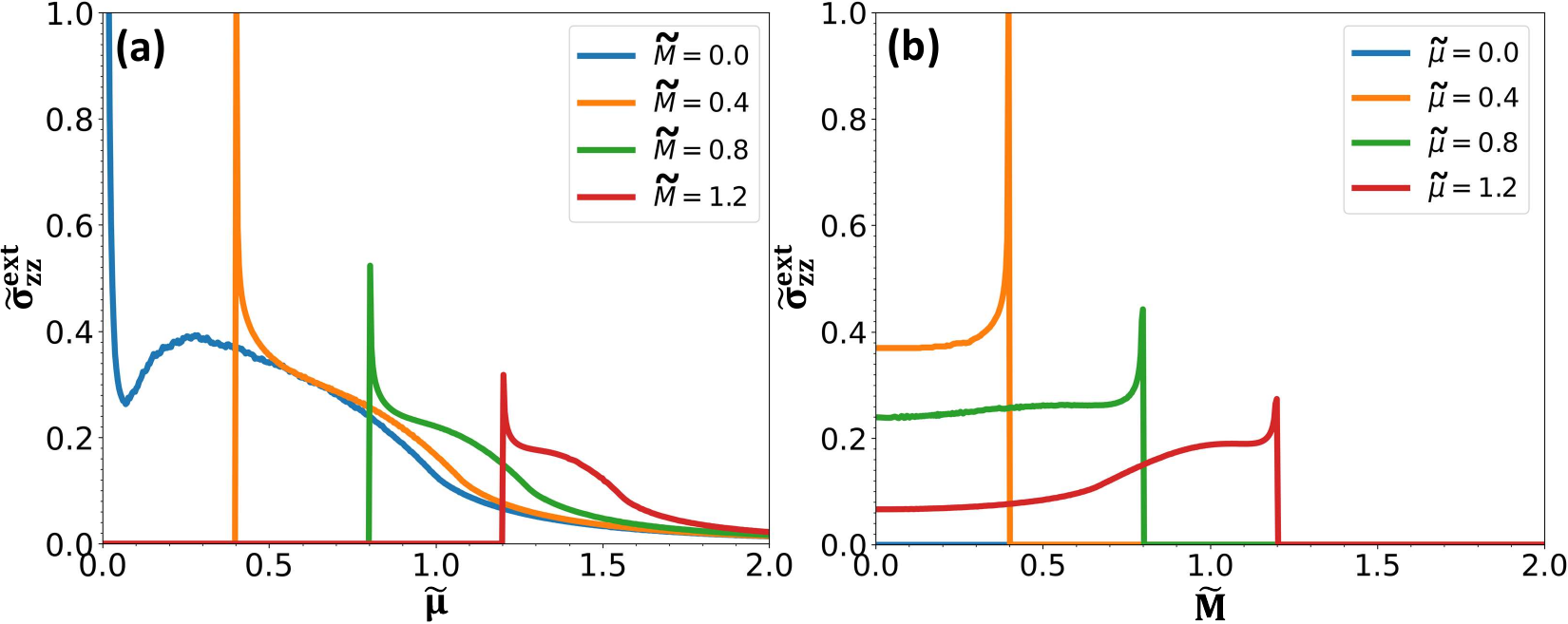}
    \caption{The extrinsic longitudinal response shows variation with (a) the chemical potential at different mass values, and (b) the mass at the different chemical potential values. Here, we set the scattering time $\tau_{ie} = 1$ ps.}
    \label{fig:2.2}
\end{figure}
\section{Results and Discussion}
\label{sec: results}

We started our discussion by numerically analyzing the longitudinal conductivity for the DNLSMs with respect to the chemical potential and gap. According to Eq.~\eqref{eqn:sigintzz}, the intrinsic conductivity is the Fermi sea contribution, as it includes the difference between the Fermi distribution function corresponding to the conduction and valence bands. This yields significant variation in the conductivity $\sigma_{zz}^{\text{int}}$, as shown in Fig.~\ref{fig:2.1}(a). To analyze further, we split the variation of the intrinsic conductivity into two different regions with reference to the energy scale $\epsilon_0 = \hbar^2 k_0^2/2m$. These regions are $\Tilde{\mu} < 1$ and $\Tilde{\mu}> 1$. 

We find that the longitudinal conductivity due to interband effects varies quadratically with the chemical potential (i.e., $\propto\Tilde{\mu}^2$) in the region  $\Tilde{\mu} < 1$ due to the higher concentration of charge carriers for the conduction. However, at higher chemical potential ($\Tilde{\mu} > 1$), the presence of the populated states saturates the conductivity. Another interesting feature is the variation of $\sigma_{zz}^{\text{int}}$ at different mass values $\Tilde{M}$. The mass term opens the gap between the conduction and valence bands, which affects the conductivity. Here, we observe that $\Tilde{\sigma}_{zz}^{\text{int}}$ decreases with a large mass term $\Tilde{M}$. A similar feature can also be seen in Fig.~\ref{fig:2.1}(b), which represents the variation of $\sigma_{zz}^{\text{int}}/\sigma_0$ as a function of $\Tilde{M}$ at distinct chemical potential values. Clearly, at zero gap value ($\Tilde{M}=0$) where Dirac nodal line contribution is also present and zero chemical potential, the intrinsic conductivity vanishes. At finite $\Tilde{\mu}$, the conductivity gives a large contribution in the absence of the gap ($\Tilde{M}=0$) and decays towards zero in the presence of a large gap due to the lack of excitations of charge carriers to the conduction band. However, on physical grounds, our numerical results for the Dirac nodal line semimetals are valid in the regime $\Tilde{\mu} \gg \Tilde{g}$.

In Fig~\ref{fig:2.2}, we plot the longitudinal extrinsic conductivity at different mass values and different values of chemical potential. Here, the conductivity yields a finite value at $\Tilde{M}\leq \Tilde{\mu}$. However, at $\Tilde{M} > \Tilde{\mu}$, i.e., on exceeding the gap over the chemical potential, the conductivity vanishes. The feature of extrinsic conductivity is explained below.

In the case of $\Tilde{M}\neq0$, there will be two regions, first when $\Tilde{M} > \Tilde{\mu}$, the system has no extrinsic conductivity contribution due to the absence of free electrons in the band gap region. However, When $\Tilde{M} \leq \Tilde{\mu}$, or when the mass touches the conduction band (i.e., is equal to the chemical potential), conduction starts, and extrinsic conductivity shows a peak due to the presence of the factor $1/(\Tilde{\mu}^2 - \Tilde{M}^2)\partial f_0/\partial \Tilde{\varepsilon}_{\bm k}$ as shown in Fig.~\ref{fig:2.2}. Here, we consider the partial derivative of the Fermi distribution function with the wave vector $\partial f_0/\partial \Tilde{\varepsilon}_{\bm k}$ at low temperature as the Dirac delta function $\delta(\Tilde{\varepsilon}_{\bm k} - \Tilde{\mu})$. Furthermore, $\Tilde{\sigma}_{zz}^{\text{ext}}$ decreases quadratically with the chemical potential.

\begin{figure}[t]
    \centering
    \includegraphics[width=5cm]{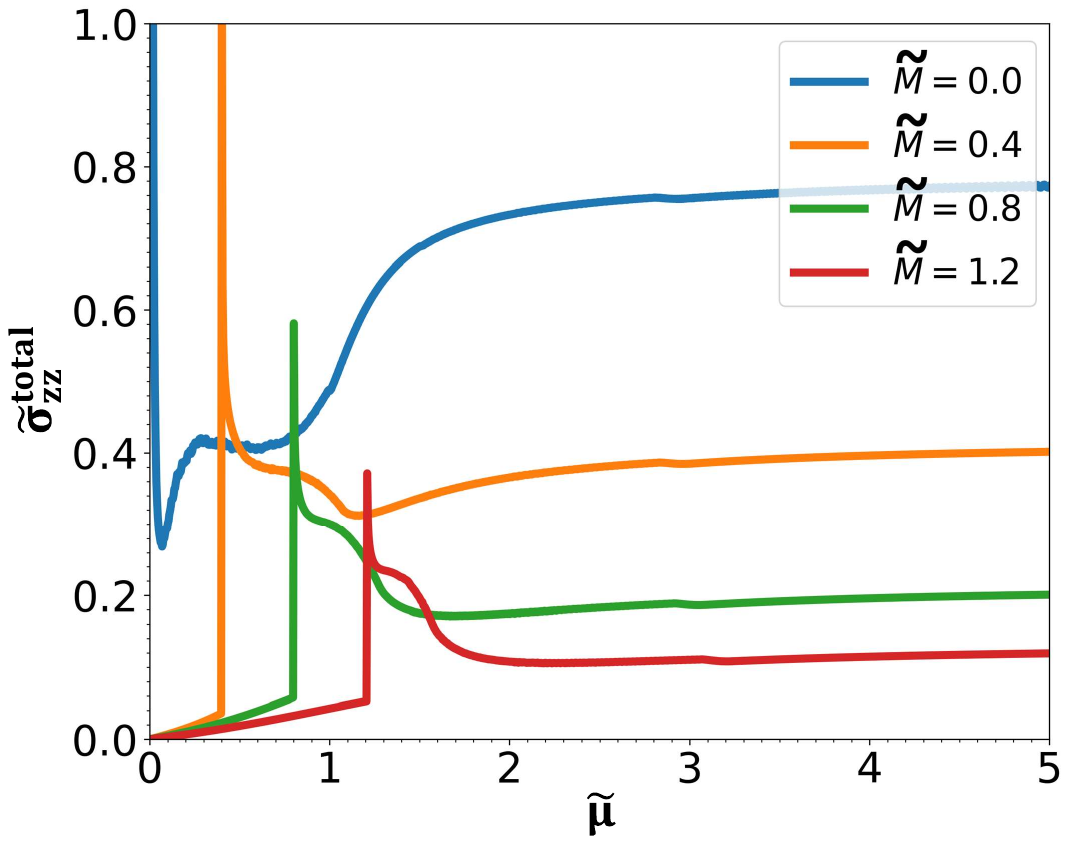}
    \label{fig:figintsig}
    \caption{Plot of the total normalized longitudinal conductivity due to interband effects as a function of the chemical potential at different mass values.}
    \label{fig:2.3}
\end{figure}

To observe the behavior of the total longitudinal conductivity due to interband effects of the DNLSMs, we plot ${\sigma}_{zz}^{\text{total}} = {\sigma}_{zz}^{\text{int}}  + {\sigma}_{zz}^{\text{ext}}$ in Fig~\ref{fig:2.3}. We find that the extrinsic part makes a major contribution at low chemical potential region $\Tilde{\mu}$ and at
high chemical potential, the total conductivity follows the intrinsic behavior and saturates. On the contrary, studies have found that the traditional Drude intraband conductivity follows a linear dependence on the chemical potential at all energy scales and at zero gap DNLSMs~\cite{barati_prb2017}.  Further, with the inclusion of a gap, the Drude conductivity is suppressed below the gap value. To compare our analysis, we find that the longitudinal conductivity, specifically comes from the extrinsic term dominates over the Drude intraband conductivity at low chemical potential.  The comparison plot for their strengths is given in Appendix~\ref{Appendix:C}. In the limit $\Tilde{g}\ll\Tilde{\mu}<1$ (or $\hbar/\tau \ll\mu < \varepsilon_0$) where our results are valid, the intraband part (Drude) of the conductivity at $\Tilde{\mu} \leq \Tilde{M}$ remains zero. However, the interband part of the conductivity gives non-zero results. Therefore the total response of the system here is purely driven by the interband contributions. Numerically by setting $\tau = 1$ps, $\hbar =  6.6 \times 10^{-16}$ eV and $\varepsilon_0 = 0.18$ eV, the limit corresponds to $0.0036 \ll \Tilde{\mu}< 1$, or an energy range of $0.66$ meV $\ll \mu <$$184$ meV. In terms of carrier density, this range spans from $10^{16}$ cm$^{-3}$ to $10^{20}$ cm$^{-3}$.
Below, we give the numerical estimation of both intrinsic and extrinsic contributions to obtain the total interband longitudinal conductivity. 

At the mass value equal to the chemical potential, $\Tilde{M}=\Tilde{\mu}=0.4$, the intrinsic conductivity yields ${\sigma}_{zz}^{\text{int}}\approx 0.04\sigma_0$ where  $\sigma_0 = \frac{e^2\gamma^2}{\hbar (2\pi)^2}$ and the extrinsic conductivity gives ${\sigma}_{zz}^{\text{ext}} \approx 1.05\sigma_0$. Hence, the relative contribution of intrinsic and extrinsic longitudinal conductivity is ${\sigma}_{zz}^{\text{ext}} \approx 26{\sigma}_{zz}^{\text{int}}$ and the total conductivity, obtained by adding the intrinsic and extrinsic contributions, is ${\sigma}_{zz}^{\space \text{total}} \approx 1.09\sigma_0$, which is mainly contributed by the extrinsic contribution. 
However, with increasing mass, the ratio between extrinsic and intrinsic longitudinal conductivity decreases. For example, at $\Tilde{M}=\Tilde{\mu}=0.8$, the ratio between the intrinsic and extrinsic conductivities becomes ${\sigma}_{zz}^{\text{ext}} \approx 8 {\sigma}_{zz}^{\text{int}}$. 
On the other hand, at $\Tilde{M}=0.4$ and $\Tilde{\mu}=1.66$, we obtain ${\sigma}_{zz}^{\text{int}} \approx 0.32\sigma_0$ and ${\sigma}_{zz}^{\text{ext}} \approx 0.03\sigma_0$, which gives the relative ratio as ${\sigma}_{zz}^{\text{int}} \approx 11{\sigma}_{zz}^{\text{ext}}$, and the total longitudinal conductivity as ${\sigma}_{zz}^{\space \text{total}} = 0.35\sigma_0$. 
Our calculations focus on the regime where $\mu \tau/\hbar \gg 1$, in which the Drude part of the conductivity dominates but the inter-band part of the conductivity is still significant. The later part can be extracted by plotting the conductivity as a function of the chemical potential $\mu$, (which can be tuned by a top gate). By extrapolating the data to $\mu = 0$, the interband contribution becomes clear.
As  $\mu \tau$ approaches 1, the inter-band contribution becomes increasingly important. In this regime, weak localization corrections must be taken into account. However, these corrections can be removed in experimental measurements by applying a magnetic field.

Although the Drude signal is dominant, extracting the interband part of conductivity gives an indication of Zitterbewegung, serving as an important probe of interband dynamics. In the case of a weak disorder (or weak localisation), the presence of interband coherence makes zitterbewegung detectable. On the other hand, in strong disorder (or strong localisation), the Bloch band picture breaks down and the Fermi surface becomes ill-defined. Further the zitterbewegung due to the absence of interband coherence is suppressed in this regime. The process of isolating the inter-band part of conductivity is somewhat analogous to weak localization experiments, where the much larger Drude conductivity dominates, but the localization effects can be observed by applying a magnetic field to extract the weak localization contribution. Similarly, Zitterbewegung is embedded within the overall conductivity, and its extraction can reveal key aspects of inter-band processes.

%

\section{Summary}
\label{sec: summary}
To summarize, we have studied the longitudinal conductivity due to interband effects of a $\mathcal{PT}$-symmetry broken three-dimensional DNLSM under a static electric field by employing the quantum kinetic approach. The total longitudinal conductivity comprises intrinsic and extrinsic interband contributions. Specifically, the intrinsic contribution stems from the off-diagonal components of the density matrix containing the field-driving terms and represents the Fermi sea effect. However, the extrinsic contribution arises from the off-diagonal scattering elements, which rely on the diagonal elements of the density matrix and represents the Fermi surface effect. 
We observe that both the intrinsic and extrinsic contributions play a significant role in enhancing the overall strength of the conductivity of the system. Furthermore, we find that
the intrinsic conductivity decreases with increasing mass terms, which enlarges the gap between conduction and valence bands. However, 
when the chemical potential nears the mass value (or touches the bottom of the conduction band), a notable peak is observed in the extrinsic conductivity. Additionally, the intrinsic contribution is prominent at a high chemical potential, whereas the extrinsic contribution dominates at a lower chemical potential. Apart from the mass and chemical potential dependence, the intrinsic and extrinsic conductivity show the scattering time dependence in different energy regimes.
Our work suggests that the presence of the disorder is pivotal, and the analysis can be extended for the varying interband scattering time by taking into account screened and unscreened Coulomb scattering potentials. In addition, similar investigations can be made for other three-dimensional materials, which will be beneficial for the future new-generation devices.

\section*{Acknowledgment}
This work is financially supported by the Science and Engineering Research Board-State University Research Excellence under project number SUR/2022/000289. 
\onecolumngrid
\appendix
\section{Form of the product of the scattering potentials} \label{Appendix:A}

The general form of the product of the scattering potential in the band basis representation is
\begin{align}
     U_{\bm{kk}'}^{nq}U_{\bm{k}'\bm{k}}^{qt}= U_0^2 \langle u_{\bm k}^{n} | u_{{\bm k}'}^{q} \rangle \langle u_{{\bm k}'}^{q} | u_{{\bm k}}^{t} \rangle,
\end{align}
where $U_0$ is the strength of the disorder potential. For the two-band DNLSM system, the product of the scattering potentials can be expressed in the form
\begin{align}
  & {U}_{\bm{kk}'}^{+-}\space{U}_{\bm{k}'\bm{k}}^{--}= \frac{U_0^2}{4}\space \bigg(1-\frac{\Tilde{M}}{\Tilde{\varepsilon}_{\bm {k}'}}\bigg) \sqrt{1-\frac{\Tilde{M}^2}{\Tilde{\varepsilon}_{\bm k}^2}} \bigg\{1-\frac{\Tilde{\varepsilon}_{\bm {k}'}+\Tilde{M}}{\Tilde{\varepsilon}_{\bm {k}'}-\Tilde{M}}
+\sqrt{\frac{\Tilde{\varepsilon}_{\bm {k}'}+\Tilde{M}}{(\Tilde{\varepsilon}_{\bm k}^2-\Tilde{M}^2)(\Tilde{\varepsilon}_{\bm {k}'}-\Tilde{M})}}(\space 2i\space\Tilde{\varepsilon}_{\bm {k}} \sin{\Gamma}+\space 2\Tilde{M}\space\cos{\Gamma})\bigg\},
\end{align}
where the angle $\Gamma=(\theta_{\bm{k'}}-\theta_{\bm{k}})$. Similarly, for the case of ${U}_{\bm{kk'}}^{++}\space{U}_{\bm{k'k}}^{+-}= U_0^2 \langle u_{\bm k}^{+} | u_{{\bm k}'}^{+} \rangle \langle u_{{\bm k}'}^{+} | u_{{\bm k}}^{-} \rangle$, we obtain
\begin{align}
  & {U}_{\bm{kk}'}^{++}\space {U}_{\bm{k}'\bm{k}}^{+-}= \frac{U_0^2}{4}\space \bigg(1+\frac{\Tilde{M}}{\Tilde{\varepsilon}_{\bm {k}'}}\bigg) \sqrt{1-\frac{\Tilde{M}^2}{\Tilde{\varepsilon}_{\bm k}^2}}\bigg\{1-\frac{\Tilde{\varepsilon}_{\bm {k}'}-\Tilde{M}}{\Tilde{\varepsilon}_{\bm {k}'}+\Tilde{M}}
  -\sqrt{\frac{\Tilde{\varepsilon}_{\bm {k}'}-\Tilde{M}}{(\Tilde{\varepsilon}_{\bm k}^2-\Tilde{M}^2)(\Tilde{\varepsilon}_{\bm {k}'}+\Tilde{M})}}(\space 2i\space\Tilde{\varepsilon}_{\bm {k}} \sin{\Gamma}+2\Tilde{M}\cos{\Gamma})\bigg\}. 
\end{align}

\section{Calculation of the conductivity component $\sigma_{ij}$} 
\label{Appendix:B}
\subsection{Intrinsic Contribution}
\label{Appendix:B1}
The intrinsic contribution to the conductivity in the $xy$ direction using Eq.~\eqref{eqn:sigma_zz} can be written as
\begin{align}
\sigma_{{xy}}^{int} = \frac{e^2}{\hbar} \sum_{n\neq p} \sum_{ \Tilde{\bm k}} \frac{F^{np} \Tilde{\omega}^{pn}}{\bigg\{\Tilde{g}+\space i\space\Tilde{\omega}^{np}\bigg\}} \mathcal{\Tilde{R}}_{\bm{k}_x}^{np}\mathcal{\Tilde{R}}_{\bm{k}_y}^{pn},
\end{align}
where $\Tilde{g}=\frac{\hbar}{\varepsilon_0 \tau_{ie}}$.

After some algebraic calculations, the real part of the conductivity in the $xy$ direction is reduced to 
\begin{align}
\label{eqn:sigintxy1}
    \sigma_{xy}^{int} = \frac{\sigma_0}{2}  \int \mathcal{\Tilde{K}}d\mathcal{\Tilde{K}}d\phi d\Tilde{k}_z \frac{\Tilde{g}F^{+-}}{\big({\Tilde{g}^2}+ 4\Tilde{\varepsilon}_{\bm k}^2\big)} \frac{\sin{2\phi}}{\Tilde{\varepsilon}_{\bm k}\big(\Tilde{\varepsilon}_{\bm k}^2-\Tilde{M}^2\big)}
    \bigg\{\big(\gamma^2 \Tilde{k}_z^2+\frac{\Tilde{M}^2 \big(\Tilde{\mathcal{K}}-1\big)^2}{\Tilde{\varepsilon}_{\bm k}^2}\bigg\},
\end{align}
where $\sigma_0 = \frac{e^2\gamma^2}{\hbar (2\pi)^2}$. The above equation vanishes on integrating over $\phi$. Similarly, the other components such as $\sigma_{yx}^{int}= \sigma_{xz}^{int} = \sigma_{yz}^{int}= \sigma_{zy}^{int} = \sigma_{zx}^{int} =0$.  

\subsection{Extrinsic Contribution}
\label{Appendix:B2}
For the calculation of the extrinsic conductivity $\sigma_{xy}^{\text{ext}}$,
\begin{align}
\label{eqn:sigma_xy}
\sigma_{xy}^{\text{ext}} = - \frac{ie}{\hbar} \sum_{n \neq p}\bigg\{\Tilde{\mathcal{R}}_{k_x}^{np}\frac{\mathcal{J}_{\bm{k}}^{np}(\mathcal{N}_E)\hbar\Tilde{\omega}^{pn}}{\Tilde{g}+i\Tilde{\omega}^{np}}\bigg\}.
\end{align}
Here, the scattering term associated with the field along the $y$ direction is
\begin{align}
\label{eqn:j_xx}
    \mathcal{J}^{+-}_{\bm{k}} [\mathcal{N}_E] &= \frac{i\pi n_i U_0^2}{2 k_0} \frac{e E_y}{\hbar^2} \sum_{k'} \frac{\sqrt{\Tilde{\varepsilon}^2_{\bm{k}'}-\Tilde{M}^2}}{\Tilde{\varepsilon}_{\bm{k}'}} \\[2ex]
    &\times \bigg\{-\tau_- \sin{\Gamma} \frac{\partial \Tilde{\varepsilon}_{\bm{k}'}^-}{\partial k'_y} \delta(\Tilde{\varepsilon}_{\bm{k}'}^- - \Tilde{\mu}) 
    \big[ \delta(\Tilde{\varepsilon}_{\bm{k}}^+ - \Tilde{\varepsilon}_{\bm{k}'}^-) + \delta(\Tilde{\varepsilon}_{\bm{k}}^- - \Tilde{\varepsilon}_{\bm{k}'}^-) \big] 
     + \tau_+ \sin{\Gamma} \frac{\partial \Tilde{\varepsilon}_{\bm{k}'}^+}{\partial k'_y} \delta(\Tilde{\varepsilon}_{\bm{k}'}^+ - \Tilde{\mu}) 
    \big[ \delta(\Tilde{\varepsilon}_{\bm{k}}^+ - \Tilde{\varepsilon}_{\bm{k}'}^+) + \delta(\Tilde{\varepsilon}_{\bm{k}}^- - \Tilde{\varepsilon}_{\bm{k}'}^+ ) \big] \bigg\}.
\end{align}
Furthermore, using the partial differentiation of the energy eigenvalue with respect to the wave vector in the $y$ direction as, $\frac{\partial{\Tilde{\varepsilon}_{\bm {k}'}^-}} {\partial{{k'}_y}}= 
\frac{\Tilde{k'}_y(\Tilde{\mathcal{K'}}-1)}{\Tilde{\mathcal{K'}}\sqrt{(\Tilde{\mathcal{K'}}-1)^2+\gamma^2 {\Tilde{k'}_z}^2+M^2}}$,
the above expression simplifies to 
\begin{align}
    &\mathcal{J}^{+-}_{\bm{k}} [\mathcal{N}_E]=0. 
\end{align}
The above equation vanishes due to the angular integration. Hence, the associated conductivity components such as $\sigma_{yx}^{\text{ext}}=0$.
Similarly, the other components due to the vanishing angular integration $\sigma_{xx}^{\text{ext}}=\sigma_{yy}^{\text{ext}}=\sigma_{xz}^{\text{ext}} = \sigma_{yz}^{\text{ext}}= \sigma_{zy}^{\text{ext}} = \sigma_{zx}^{\text{ext}} =0$. 
\newpage
\section{Comparison between Drude (intraband effect) and conductivity due to interband effects}\label{Appendix:C}
\begin{figure}[htp]
    \centering
    \includegraphics[width=6.5cm]{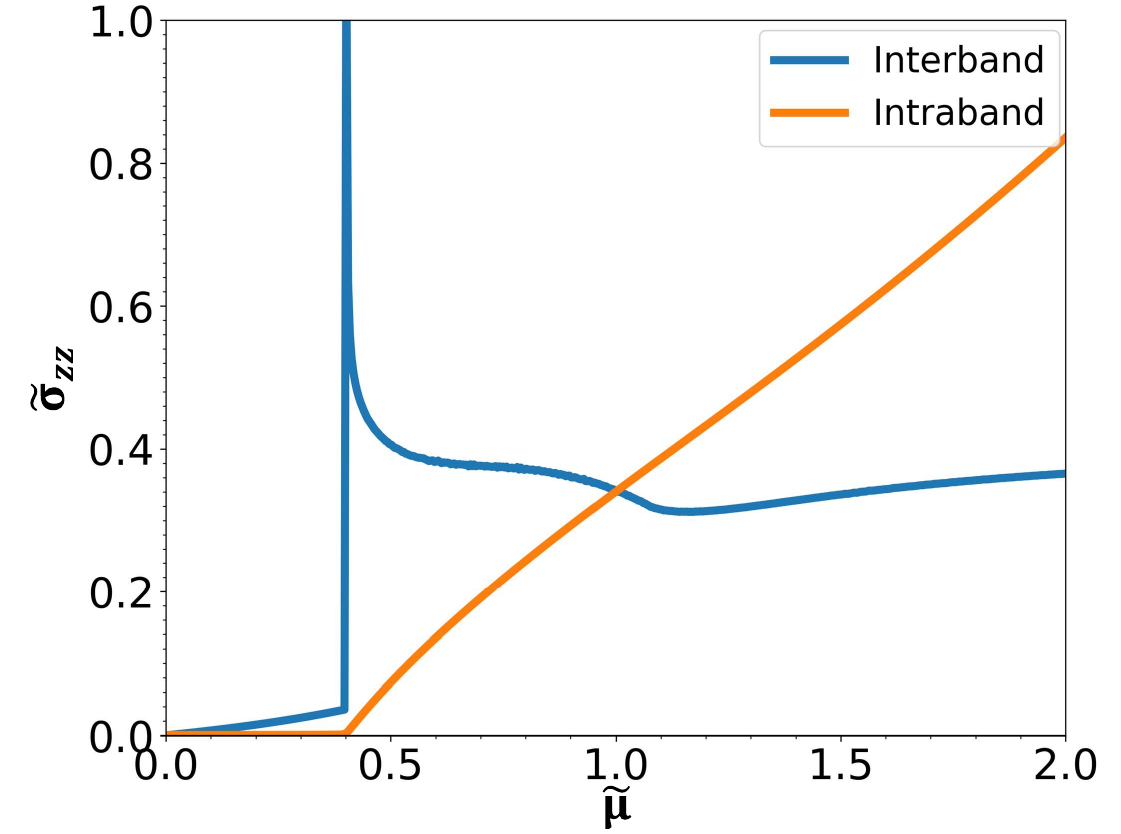}
    \caption{The plot shows the intraband and interband parts of conductivity of the DNLSM for the mass value $\Tilde{M}=0.4$. Here, the interband part at low chemical potential dominates, whereas, at high chemical potential the intraband part takes the lead.}
    \label{fig:3.0}
\end{figure}
\twocolumngrid
\bibliography{Ref}

\end{document}